# Modeling and characterizing single-walled carbon nanotubes by pressure probe

**Ali Nasir Imtani and V. K. Jindal***

Department of Physics, Panjab University, Chandigarh-160014, India

## Abstract

We compare the behavior of bond lengths, cross sectional shape and bulk modulus in equilibrium structure at ambient conditions and under hydrostatic pressure of all the three kinds of uncapped single walled carbon nanotubes. Results of our numerical calculations show that two bond lengths completely describe the structure of achiral SWNT whereas only one bond length is required to determine structure of chiral SWNT. In armchair tubes, one bond length is found to be larger than that of graphitic value while in zigzag tubes one bond length has a constant value. These bond lengths are very sensitive to tube radius. In chiral tubes, the value of bond length is found to depend on the chirality and slightly on the tube radius. Different responses of these bond lengths are found on application of pressure. At some critical pressure, both bond lengths become equal to each other in achiral tubes. An analysis regarding the cross sectional shape of the nanotubes and its pressure dependence has also been done. The shape transition, from circular to oval shape takes place. At this transition, the behavior of bond lengths is found different and dependent on the chirality of the tubes. Chiral tubes with chiral angle which is mid way between zigzag and armchair tubes are found to have most prominent effects of chirality. Thus we demonstrate that pressure is a useful probe to characterize various kinds of carbon nanotubes.

## I Introduction

Carbon nanotubes rank among the most exciting new developments in modern science and engineering. Since carbon nanotubes[1,2] were first discovered by Iijima[3], the past 10 years witnessed significant progress in both carbon nanotubes synthesis and investigations on their electrical, mechanical and chemical properties. This has been largely driven by the exciting science involved and numerous proposed applications of carbon nanotubes due to their unique electronic properties and nanometer sizes.

In this paper, we present a comparison of the behavior of bond lengths and some transition and critical pressures at which interesting changes in the behavior of various single-walled carbon nanotubes (SWNT) take place. SWNT can be viewed as a sheet of graphite rolled into seamless cylinders with nanometer scale diameters and micrometer scale lengths [1].

* Author with whom correspondence be made, e-mail: jindal@pu.ac.in

The difference between three types of single-walled carbon nanotubes is that the direction of bond lengths with the tube axis is different in the three cases. There are two different directions of the bond lengths in achiral tubes while three different directions in the structure of chiral tubes. In armchair tubes, the direction of bond length $b_1$ is perpendicular to the tube axis, and parallel to the tube axis for zigzag tubes. The other bond length $b_2$ forms an angle with tube axis in achiral tubes (see Fig. 1(a) and Fig 1(b)). In chiral tubes, all three bond lengths form an angle with tube axis (see Fig.1(c). For this reason, one expects that these bond lengths can have different values.

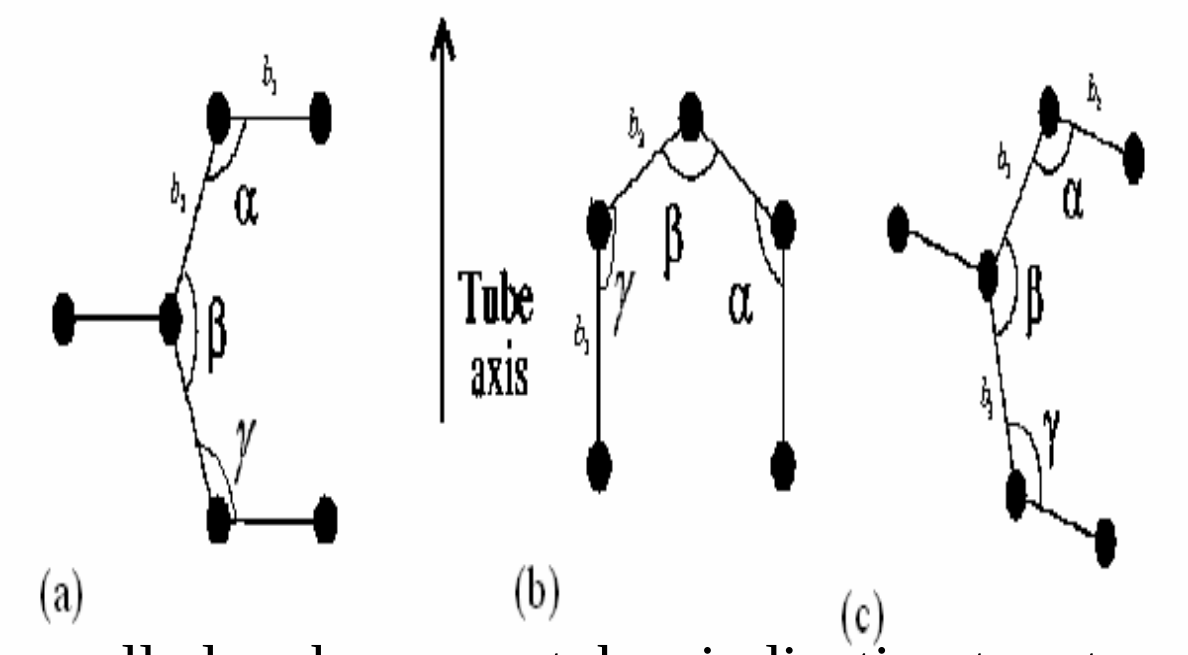

Figure 1: A part of single-walled carbon nanotubes indicating two types of C-C bond lengths; these are labeled as $b_1$ and $b_2$ for armchair tubes (a), zigzag tubes (b) and three type for C-C bond length for chiral tubes (c).

Bond lengths for some tubes have already been studied using ab initio calculations [4-7]. Many experimental and theoretical studies have also been done on bundles as well as isolated single-wall nanotubes under pressure [8-13]. Those studies reveal that there exists a transition pressure where either the tubes collapse [8] or the vibrational modes change [9] or loss of reversible deformation [11] takes place. The transition pressure has been found to depend on the diameter of the nanotubes and not on its chirality [8]. All those studies have investigated the value of the transition pressure of armchair tubes, that too generally for (10,10) tubes. No details of the behavior of the bond length variation under pressure have been made.

**II Theoretical procedure**

In our previous studies [14-16], we described our numerical procedure which is applied to obtain results of the variation of bond lengths in SWNTs from that of graphite sheet at ambient pressure. We also presented the numerical procedure to describe the behavior of the bond lengths under hydrostatic pressure assuming that the cross section retains circular shape and then the cross section of tubes collapsed to oval shape. In the following, we compare results of the three types of tubes.

## II Results and Discussion

### (1) Bond lengths as a function of tube radius and chirality

Normalized bond lengths as a function of tube radius have been plotted in Fig.2 for three types of SWNTs. We observe that two bond lengths completely describe the structure of achiral tubes while one bond length for chiral tubes. In the structure of armchair tubes, the bond length, $b_1$, has value larger than that of the graphitic value (Fig.2(a)) while this bond length has a constant value equal to graphitic value in the structure of zigzag SWNTs(Fig.2(b)). The other one $b_2$ has value smaller than the graphitic value in armchair and while larger than that of graphitic value in the structure of zigzag tubes. These bond lengths are very sensitive to the tube radius. As the tube radius becomes large, the bond lengths approach to that of graphite.

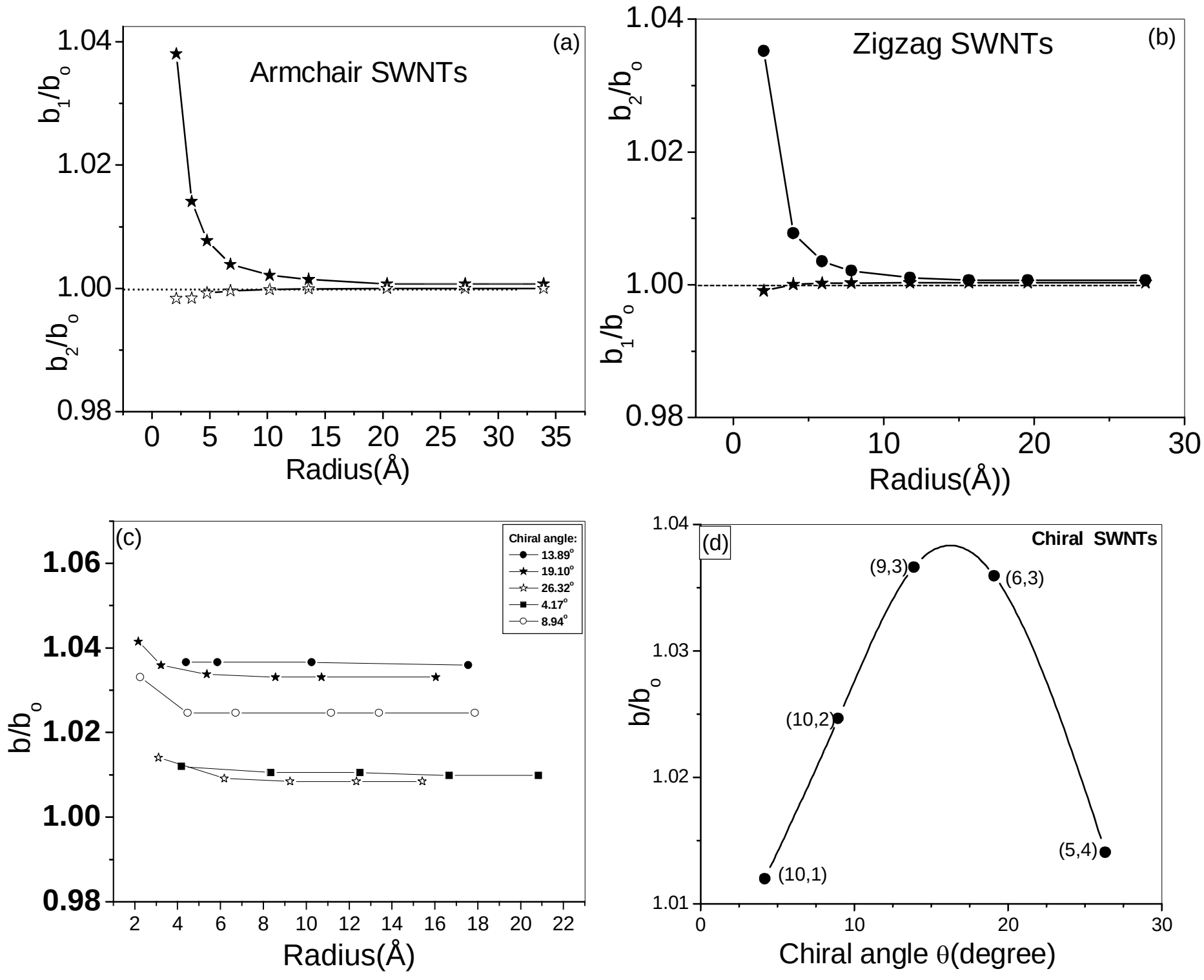

Figure 2: Normalized bond lengths as a function of the tube radius for (a) armchair tubes, (b)zigzag tubes and (c) chiral tubes at five different chiral angles. (c) Normalized bond length versus chiral angle for tubes having approximately the same radius.

Results of chiral tubes indicate that one bond length completely describes their structures. The bond length in chiral tubes of the same chiral angle is quite independent of the radius of tubes as shown in Fig. 2(c). Only for very small radii tubes, there is a slight effect of tube radius on this bond length. However, chiral angle ( $\text{chirality} = m/n$ ) is found to have a strong effect on the value of this bond length(see Fig. 2(d)). The smaller value of the bond length lies near the chiral angle of armchair tubes (chiral angle=30°) and zigzag tubes (chiral angle=0°) while the larger deviation from that of graphitic value occurs in critical chiral angle 13.892 °, in the mid region between chiral angle of zigzag and armchair chiral angles.

The work of Sanchez-Portal et al.[4] reports results of two bond lengths in the structure of armchair tubes. They found that both bond lengths in armchair tubes have values larger than that in graphite. Lin-Hiu et al.[5] also reported results of two bond lengths in the structure of armchair tubes, in which one bond length elongates as compared to that in graphite whereas the other shrinks but for very small radii tubes this tendency reverses. For the structure of zigzag tubes, Lin-Hiu et al. and Gulseren et al.[6] also found two bond lengths in this structure where $b_1$ was smaller than of the graphite value while $b_2$ was larger than it. Our results of the bond length for the structure of chiral tube disagree with the calculations done by Jiang et al.[7]. It is found that there are three unequal bond lengths for (4,2) tubes. The values of these bond lengths are equal to 1.460 Å, 1.467 Å and 1.455 Å and for (9,3) tubes, those bond lengths become 1.453 Å , 1.454 Å and 1.451 Å. In our calculations of the bond length for (9,3) and (4,2) tubes, the values of the bond length are found equal to 1.472 Å and 1.479 Å, respectively. Notice that these two tubes having different chirality and lie in the chiral angles 13.89° and 19.10°, respectively. The strong effect of the chirality appears in these chiral angle tubes.

**(2) Bond length under hydrostatic pressure**

Here in this section, we compare the variation of the bond lengths under hydrostatic pressure of three types of SWNTs in two cases involving circular and elliptical cross sectional shapes.

**(A) Circular cross section** (AB curve)

Fig.3 shows the results of our calculations of the bond lengths at various pressures for achiral and chiral SWNT's. AB curve represents a variation of two bond lengths under pressure for achiral tubes assuming that the tubes remain in circular cross section. For armchair SWNTs, we observe that the bond lengths $b_1$ and $b_2$ decrease under pressure ( Fig.3(a)). The larger bond length $b_1$ decreases faster with pressure as compared to the smaller bond length $b_2$. At some critical values of pressure ($P_c$), both bond lengths become equal to each other ($b_1 = b_2 = b_c$). $P_c$ and corresponding critical bond length ($b_c$) are dependent on tube radius. Above this critical pressure, the smaller bond length changes to become larger bond length and vice versa. Fig.3 (b), shows the results of zigzag tubes. We observe that the pressure affects only one bond length $b_2$ while the bond length $b_1$ remains unchanged. At $P = P_c$, the larger bond length $b_2$ becomes equal to the constant bond length. Above $P_c$, $b_2$ continues to compress. In chiral tubes, the bond length continues to compress under hydrostatic pressure (Fig. 3(c)). We have also calculated the bulk modulus at zero pressure of different radius and different chirality tubes. Fig.3(d) shows bulk modulus with chiral angles for tubes having approximately the same radius. Tubes of chiral angles far from zigzag and armchair chiral angle have lower value of bulk modulus.

**(B) Elliptical Cross section (CD curve)**

At shape transition, from circular to oval cross section, the behavior of bond lengths at transition pressure is quite peculiar. This transition has also been plotted in Fig. 3. In Fig.3, CD curve represents a variation of two bond lengths under applied pressure for achiral tubes. Each point on CD represents different value of the elliptical aspect ratio $b_e / a_e$, where $b_e$ and $a_e$ are the shorter and longer radius in elliptical cross section. Results show that at transition pressure, the larger bond length $b_1$ expands against pressure to take value closer to the value at zero pressure. The other bond length $b_2$ continues to compress under pressure. At pressures above transition pressure, the bond length $b_1$ remains practically unchanged although the elliptical aspect ratio changes as can be noticed from almost flat CD curve above transition pressures. Chiral tubes having different radii of the same chirality and those having approximately the same radius but different chiralities are chosen. In Fig. 3(c), we have plotted the bond length as a function of pressure for chiral tubes with different radius and different chirality. From this figure, we observe that at transition pressure (point C) the bond length reduces to have nearly the same value of bond length for all chiral tubes with different radius and different chirality. Our results of the transition pressure of (10,10) tubes($P_T$ =2.26 GPa) are in agreement with a recent experimental value ($P_T$ =2.5 GPa) [13]. The transition pressure is also found to have value lower in case of chiral tubes when compared with that in achiral tubes (Fig. 3(e)).

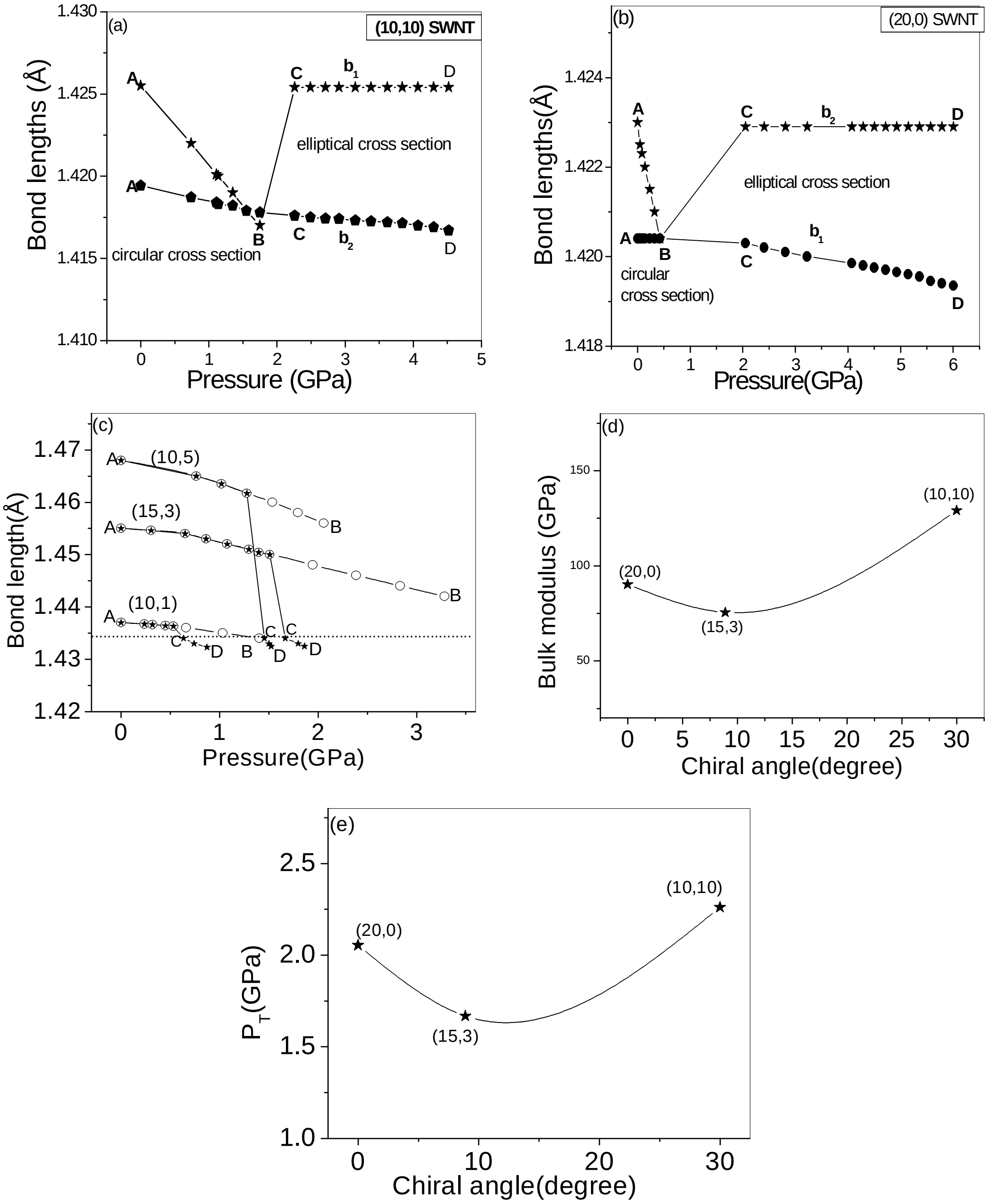

Figure 3: Bond lengths as a function of applied pressure for (a) armchair (10,10), (b) zigzag (20,0) tubes and (c) chiral (10,1), (15,3) and (10,5) tubes. AB curves correspond to circular and CD to elliptical cross section with different aspect ratio $b_e / a_e$. At transition pressure, all chiral tubes have approximately the same value of the bond length (dotted line in 3(c)). Also in (d) and (e) , bulk modulus at zero pressure and transition pressure $P_T$ versus chiral angle of tubes having approximately the same radius of different chirality are shown.

**III Summary and conclusions**

In this paper, we compared the results of the bond lengths for three types of SWNTs. These results are obtained by applying our procedure based on the helical and rotational symmetries[17] to generate the atomic coordinates and then minimized the energy through Tersoff potential [14-16,18]. Two bond lengths are found to completely describe the structure of achiral SWNTs while only one bond length is required to determine the structure of chiral SWNTs. For achiral tubes, one bond length is larger than that of graphite value. The other bond length has value smaller than graphitic value in armchair tubes while in zigzag tubes this bond length has a constant value equal to graphitic value. These bond lengths in achiral tubes are found sensitive to the tube radius. In the structure of chiral tubes, the bond length is found to strongly depend on the chiral angle; with hardly any dependence on the tube radius.

These bond lengths have different responses with application of hydrostatic pressure depending on the chirality and the shape of cross section of the tubes. Below a transition pressure when the tubes have circular cross section, these bond lengths become equal at some critical pressure in achiral tubes. At transition pressure, cross sectional shape changes from circular to oval shape and one bond length expands to take value closer to value at zero pressure. The second bond length continues to compress in achiral tubes. In chiral tubes, the bond length has approximately the same value at transition pressure. The cross section of chiral tubes requires lower transition pressure to collapse compared to achiral tubes. Similarly chiral tubes have lower values of bulk modulus in comparison to achiral tubes.

Therefore, we propose that pressure is a very useful probe to characterize the type and radius of SWNT. The critical and transition pressures suggested here motivate experimental activity aimed to measure vibrational characteristics of SWNT or their bundles under hydrostatic pressure. It is interesting to note that some effort in this direction has already been made [13, 19] and transition pressure obtained seems be close to predicted here for (10,10) tubes. However, much more extensive experimental work on a variety of tubes is required to be made.